\documentclass[journal,comsoc,10pt]{IEEEtran}

\usepackage[T1]{fontenc}

\usepackage{makecell}
\usepackage{cite}
\usepackage[colorlinks,
linkcolor=red,
anchorcolor=blue,
citecolor=green
]{hyperref}

\ifCLASSINFOpdf
\else
\fi
\usepackage{amsmath}
\usepackage{diagbox}

\usepackage{ amssymb }
\usepackage{amsfonts}
\usepackage{caption}
\usepackage{graphicx}
\usepackage{lettrine}
\usepackage{epstopdf}
\usepackage{textcomp,booktabs}
\usepackage[usenames,dvipsnames]{color}
\usepackage{colortbl}
\definecolor{mygray}{gray}{.9}
\definecolor{mypink}{rgb}{.99,.91,.95}
\definecolor{mycyan}{cmyk}{.3,0,0,0}
\usepackage{pifont}
\usepackage{subfigure}
\usepackage{multirow}
\usepackage{url}
\usepackage{color}
\usepackage{threeparttable}
\usepackage{setspace}
\usepackage{lineno}
\usepackage{tabularx}
\usepackage[dvipsnames,usenames]{color}

\usepackage{bm}

\usepackage{algorithm} 
\usepackage{algorithmic} 

\usepackage{dblfloatfix}

\usepackage[dvipsnames,usenames]{color}
\usepackage[usenames,dvipsnames]{color}

\definecolor{light-gray}{gray}{0.90}
\begin{document}

	\title{AgentComm: Semantic Communication for Embodied Agents}

	\author{Peiwen Jiang,~\IEEEmembership{Member,~IEEE,} Yushuo Feng,  ~\IEEEmembership{Graduate Student Member,~IEEE,} Jiajia Guo,~\IEEEmembership{Member,~IEEE,} Chao-Kai Wen,~\IEEEmembership{Fellow,~IEEE,} and Shi Jin,~\IEEEmembership{Fellow,~IEEE} 
			\thanks{P. Jiang, Y. Feng, J. Guo and S. Jin is with the School
				of Information Science and Engineering, Southeast University, Nanjing
				210096, China (e-mail: peiwenjiang@seu.edu.cn, yushuofeng@seu.edu.cn, jiajiaguo@seu.edu.cn, jinshi@seu.edu.cn).}
			\thanks{C.-K. Wen is with the Institute of Communications Engineering, National
				Sun Yat-sen University, Kaohsiung 80424, Taiwan (e-mail: chaokai.wen@mail.nsysu.edu.tw).}}
	
	\maketitle
	\pagestyle{empty}  
	\thispagestyle{empty} 

\begin{abstract}
The increasing deployment of agentic artificial intelligence (AI) systems has intensified the demand for efficient agent to agent communication, particularly over bandwidth limited wireless links. In embodied AI applications, agents must exchange task related information under strict latency and reliability constraints. Existing agent communication methods primarily focus on connectivity and protocol efficiency, but lack effective mechanisms to reduce physical layer transmission overhead while preserving task semantics.
To address this challenge, this paper proposes a semantic agent communication framework that reduces communication overhead while maintaining task performance and shared understanding among agents. An LLM based semantic processor is first introduced to reorganize and condense agent generated messages by extracting task relevant semantic content. To cope with information loss introduced by aggressive message reduction, an importance-aware semantic transmission strategy is developed, which adaptively protects semantic components according to their task importance. Furthermore, a task specific knowledge base is incorporated as long term semantic memory to support recurring tasks and further reduce bandwidth consumption with minimal performance degradation. Experimental results and ablation studies demonstrate that the proposed framework achieves nearly 50\% bandwidth reduction with negligible loss in task completion performance compared to conventional transmission schemes.

\end{abstract}


\section{Introduction}

\IEEEPARstart{E}{mbodied} artificial intelligence (AI) bridges AI with real world physical systems by enabling autonomous perception, learning, and action in complex environments. A rapidly evolving direction is agentic embodied AI \cite{feng2025multi}, where multiple AI agents proactively make decisions and collaboratively execute tasks with minimal human intervention. In practical deployments such as smart city traffic management, autonomous driving, and emergency response, these agents must exchange task related information under strict latency and reliability constraints over wireless links \cite{lian2017game,zhang2025embodied}. Recent advances in large language models (LLMs) further enhance agents' reasoning and planning capabilities, enabling flexible coordination under decentralized control \cite{zhang2023building}. Consequently, communication efficiency has become a key factor determining the scalability, responsiveness, and robustness of embodied AI systems. However, existing agent communication mechanisms remain inefficient from a communication system perspective.

From the above description, a critical question arises: Why do LLM-driven agents make the existing communication inefficiency problem even more severe in the physical layer?   Agent generated messages, particularly those produced by LLM-driven planning and reasoning, often contain substantial redundancy and task irrelevant information. When transmitted over bandwidth limited and noisy wireless channels, such messages introduce excessive physical layer communication overhead, which becomes a major performance bottleneck. This issue is further exacerbated in dynamic environments with constrained computational and networking resources, where inefficient interactions significantly degrade task execution and system responsiveness \cite{duan2022survey,fung2025embodied}. Therefore, reducing physical layer transmission overhead while preserving task semantics and coordination effectiveness is a fundamental challenge for agentic embodied AI systems. 

Agent to agent (A2A) communication serves as a critical enabler for cooperative planning and decentralized robotic systems, relying on either internal message passing or wireless links. Emergent communication studies \cite{lazaridou2020emergent} reveal how symbolic representations acquire meaning through interaction, clarifying the functional role of language in machine to machine and human to machine communication. Meanwhile, reliable communication over noisy channels has motivated joint learning and communication designs that explicitly account for channel impairments \cite{tung2021effective}. As a result, establishing an efficient and semantic-aware communication framework for agentic applications has become an urgent research challenge \cite{xu2024unleashing}.

\begin{figure*}
	\centering
	{\includegraphics[width=0.99\linewidth]{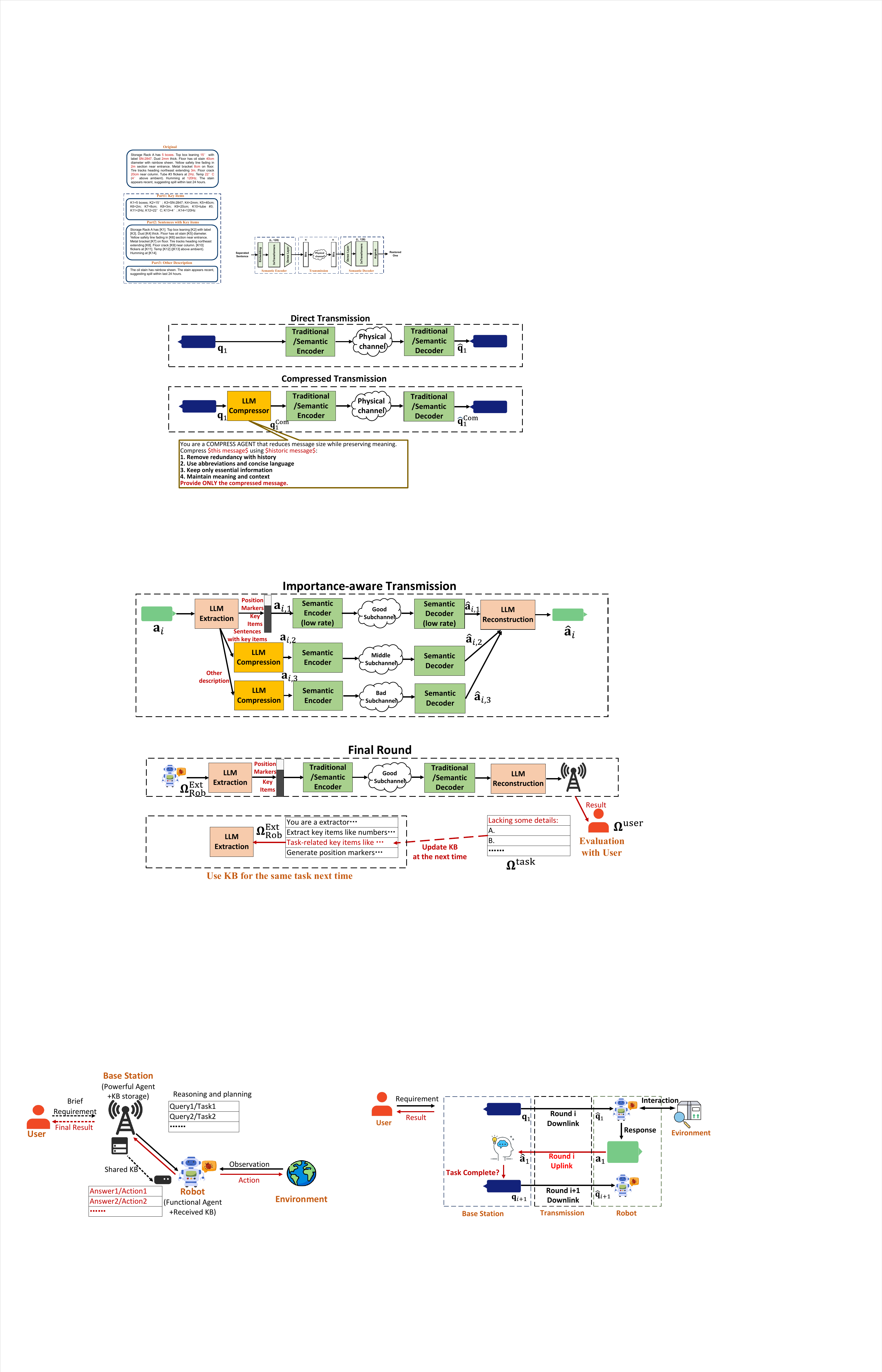}}
	\caption{Overview of the semantic agent communication system for embodied AI over wireless links. The user provides task requirements at the beginning, while a powerful BS agent collaborates with the robot agent to execute the task.}
	\label{sys}
\end{figure*}

To address the inefficiency of agent generated message transmission, semantic communication \cite{gunduz2022beyond,zhang2024intellicise} has emerged as a promising paradigm. Unlike conventional schemes that prioritize symbol level accuracy, semantic communication focuses on transmitting task relevant and meaning preserving information. This objective aligns naturally with agentic embodied AI, where communication primarily supports coordination, task understanding, and collaborative decision making. Prior studies demonstrate that semantic methods are effective for complex message content \cite{10841385,zhang2022unified,10646587} and challenging wireless environments \cite{jiang2025semantic,10896580,yi2022semantic}. Consequently, semantic transmission has attracted increasing attention as a potential solution for agent communication \cite{yu2025semantic,charalambous2025toward}. 

Although semantic communication provides a principled framework, its realization in agentic systems cannot  be separated from joint design with modern agents. For example, advances in agentic AI for communication systems \cite{jiang2025from,peyrard2024agentic,guo2025large,guo2025prompt} show that LLM powered agents can naturally function as semantic processors, capable of generating and manipulating high level semantic representations. By exchanging compact semantic tokens instead of verbose raw messages, such agents implicitly revise the conventional end-to-end encoder-decoder paradigm through shared world knowledge embedded in large models. These developments suggest that designing a unique agent for semantic processing also offers a practical and scalable pathway for implementing semantic communication in multi-agent wireless systems.

We consider a semantic agent communication scenario for embodied AI \cite{liu2021robotic,duan2022survey}, where a functional robot agent collaborates with a powerful base station (BS) agent over bandwidth limited wireless links to accomplish a task, as illustrated in Fig.~\ref{sys}. Communication focuses on task relevant semantic information rather than raw sensory data or low level control signals. The BS acts as both a secure storage unit and a powerful agent that cooperates with the robot to autonomously complete tasks, while the user provides task requirements only at the beginning and receives the final outcome upon completion. This work focuses on reducing transmission overhead within a single task, where only a limited number of BS robot communication rounds are required.

Specifically, a transmission agent is designed to compress agent generated messages by eliminating semantic redundancy. A semantic encoder-decoder is then employed to mitigate wireless channel impairments. To enhance reliability, an importance-aware transmission strategy is introduced, providing unequal protection to semantic components according to their task importance. In addition, a task specific knowledge base (KB) is maintained at the BS to support recurring tasks and compensate for semantic information loss caused by message compression. Ablation studies demonstrate that the proposed design effectively reduces transmission overhead and highlights the potential of multi agent semantic communication for embodied AI systems. 

The main contributions of this work are summarized as follows:
\begin{itemize}

\item \textbf{LLM based semantic processor for agent communication:}
We propose an LLM based semantic processor that serves as an explicit semantic abstraction layer prior to physical layer transmission in agentic embodied AI systems. By reorganizing and condensing task relevant semantic content using shared world knowledge embedded in LLMs, the proposed processor directly generate an explicit semantic abstract before physical transmission  without  training. This design significantly reduces transmission overhead while preserving task-level semantics.

\item \textbf{Importance-aware semantic transmission with flexible encoder-decoder design:}
Building upon the semantic processor, we develop an importance-aware semantic transmission strategy that jointly considers semantic relevance and physical layer reliability. Semantic message components with different task importance levels are mapped to distinct encoder-decoder configurations, enabling unequal semantic protection and improving communication efficiency without sacrificing task execution robustness.

\item \textbf{Task specific KB as long term semantic memory:}
We introduce a task specific KB that functions as long term semantic memory for recurring embodied AI tasks. By storing task-relevant semantic patterns, user preferences, and environment related information, the KB supports semantic reconstruction under aggressive message compression and effectively bridges the performance gap between compressed and full message transmission.

\end{itemize}

The remainder of this paper is organized as follows. Section \uppercase\expandafter{\romannumeral2} describes the system model and multi agent communication process. Section \uppercase\expandafter{\romannumeral3} presents the proposed framework. Section \uppercase\expandafter{\romannumeral4} evaluates the performance of the proposed transmission agent and semantic encoder-decoder. Section \uppercase\expandafter{\romannumeral5} concludes the paper.

\begin{figure*}
	\centering
	{\includegraphics[width=0.99\linewidth]{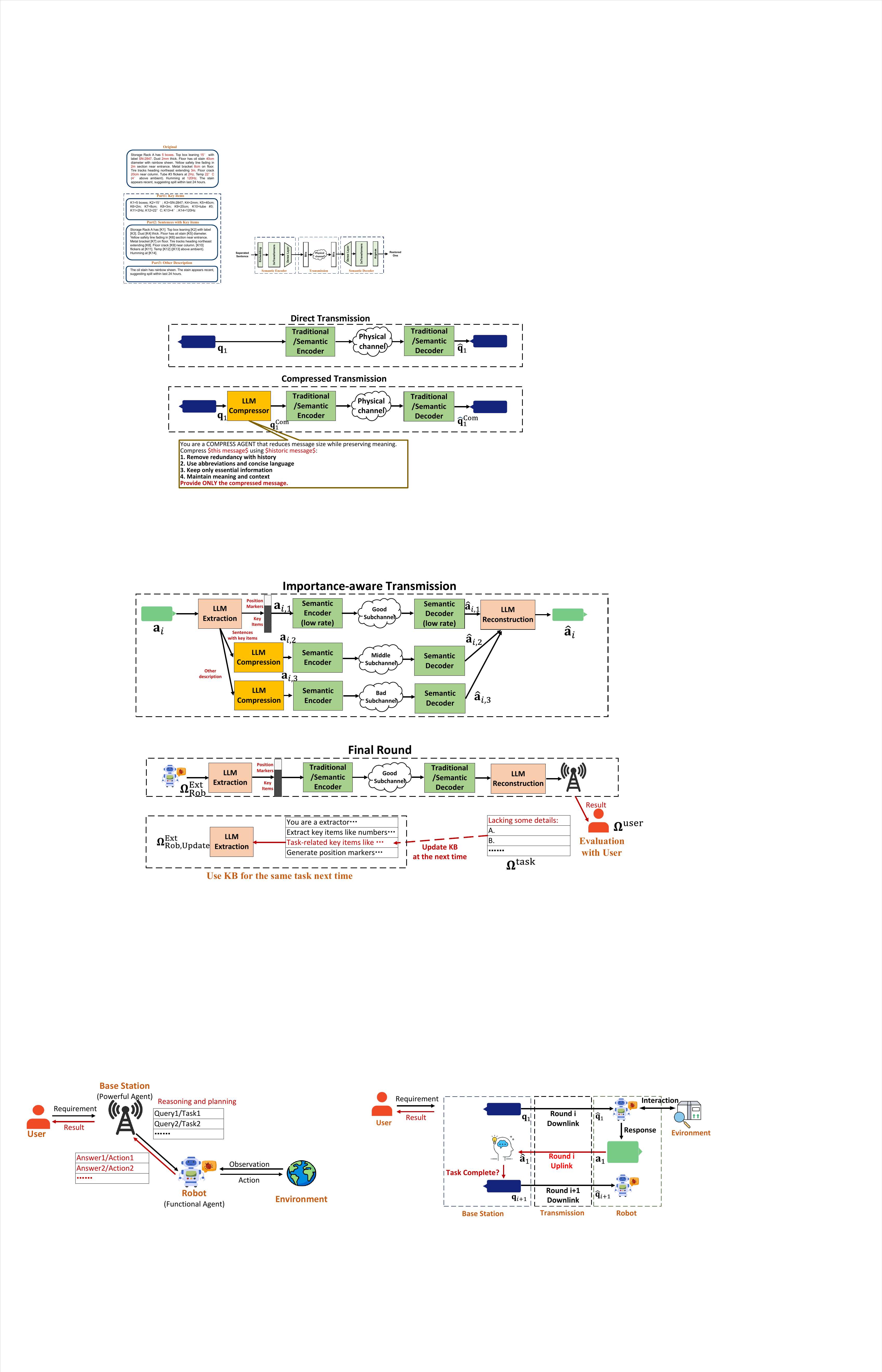}}
	\caption{Overview of the proposed communication process, where multi-round transmissions are applied.}
	\label{Overall}
\end{figure*}

\section{System Model}
\label{SystemModel}
In this section, we present the system model for semantic agent communication in embodied AI systems. We first describe the roles and functionalities of AI agents deployed at different nodes, with emphasis on their collaboration under wireless communication constraints. The A2A communication process over bandwidth limited wireless links is then introduced, followed by a description of the underlying physical layer transmission model. Finally, we discuss the motivations and challenges of integrating semantic encoder-decoders with agent based processing.

\subsection{Agents for Embodied AI}
AI agents are autonomous software entities designed to perceive their environment, make decisions, and take actions to achieve specific goals \cite{zhang2024aflow,besta2024graph}.  The differences between traditional AI and agentic AI systems are summarized as follows:
\begin{enumerate}
    \item \textbf{Traditional AI} performs well defined tasks such as recognition, classification, and recommendation. These systems operate reactively by mapping inputs to outputs and cannot autonomously set goals, execute action sequences, or adapt beyond their training. Even when equipped with LLMs, traditional systems still rely on explicit prompts and generate outputs in a step by step manner. As a result, they cannot independently decompose complex problems and require substantial human intervention for complex tasks.
    
    \item \textbf{Agentic AI} operates autonomously toward predefined goals without continuous human supervision \cite{hong2023metagpt,qian2024chatdev}. These systems can plan multi-step processes, utilize tools, gather information, and iteratively refine their strategies based on feedback. Beyond question answering, agentic AI is widely adopted for tasks such as iterative code development, debugging, and project management through coordinated task execution. This paradigm represents a shift from reactive behavior to proactive goal oriented problem solving.
\end{enumerate}

At their core, the collaboration of the AI agents can fully utilize the capabilities of different nodes. Otherwise, the complex tasks can exceed the capabilities of a single agent through requiring diverse skills, knowledge, or perspectives. Through information sharing and coordinated decision making, agents can identify errors, compensate for missing knowledge, and produce more robust solutions than isolated agents. This collaborative behavior mirrors human organizational structures and enables AI systems to scale their problem solving capabilities for increasingly complex tasks. Recently, AI agents empowered by LLMs exhibit enhanced flexibility and usability by enabling natural language understanding and reasoning \cite{ferrag2025llm}. Such agents provide improved generalization, common sense reasoning, and robustness when dealing with ambiguous or unfamiliar situations.

Taking into account the requirements of the embodied applications,  AI systems play important roles in physical robots or devices that interact with the real world \cite{sumers2023cognitive}.   Unlike purely software based systems, embodied AI must operate under physical constraints, process information in real time, and handle multi-modal inputs such as vision, audio, and tactile signals simultaneously. For example, the functional agents in the robots can process different information from their surroundings  through their own sensors or data inputs. However, the single robot usually cannot employ planning and reasoning mechanisms to determine appropriate actions for the entire system with different devices because a selected device may not have access to complete user related knowledge. This limitation commonly arises due to the following reasons: 
\begin{enumerate}  
    \item \textbf{Limited memory:} Mobile devices typically lack sufficient memory to retain long term contextual information, preventing full utilization of user background knowledge. Consequently, users are required to interact with the robot through multiple communication rounds to complete complex tasks.
    
    \item \textbf{Personal privacy:} Privacy constraints restrict robots, especially public devices, from collecting sensitive user information such as location or personal habits. Such information could otherwise be inferred by observing robot behavior during interactions with different users.
\end{enumerate}

From the above description, the communication is essential for AI agents to receive user instructions, coordinate with other agents or systems, and request clarification when handling complex tasks. The wireless links for the AI agents are becoming the hotpot for the flexible connections\cite{zou2023wireless,chen2024enabling}.

\subsection{Goal-Driven Communication from Agent Collaboration}
As shown in Fig.~\ref{sys}, we consider a BS serves as a central entity that stores user profiles, interaction histories, and cloud based knowledge resources. When a user submits a task request, the BS enriches it with relevant contextual information before forwarding it to the robot. The BS also decomposes tasks into multiple steps and validates the robot responses before delivering the final results to the user. This architecture alleviates memory constraints at the robot while enabling efficient wireless communication and consistent knowledge sharing across multiple devices.

When a task request is issued, the BS selects a specific robot agent to execute the task. The corresponding queries and responses are transmitted over a wireless channel. In this work, the physical transmission of queries and responses is modeled using an orthogonal frequency division multiplexing (OFDM) system, which is widely adopted in modern wireless communication systems. Specifically, each  message, including a query $\mathbf{q}_i$ or a response $\mathbf{a}_i$, is encoded and mapped onto parallel subcarriers. After the modulation and subcarrier mapping, an inverse fast Fourier transform generates the time domain signal, followed by cyclic prefix insertion to mitigate inter symbol interference. Pilot symbols are inserted for channel estimation, and edge subcarriers are nulled to satisfy spectral constraints. The number of the effective subcarriers are denoted as $K$.

In OFDM systems, each subcarrier experiences distinct channel conditions due to frequency-selective fading. The effective signal-to-noise ratio (ESNR) is defined as the equivalent additive white Gaussian noise (AWGN) SNR that achieves the same block error rate as the actual fading channel with non-uniform subcarrier SNRs. For $K$ active subcarriers with per-subcarrier SNRs ($SNR_k$), the effective SNR is computed as
\begin{equation}
    -\beta \log \left( \frac{1}{K} \sum_{i=1}^{K} e^{-\text{SNR}_k / \beta} \right),\label{ESNR}
\end{equation}
where $\beta > 0$ is a parameter whose value depends on the used modulation and coding scheme (MCS) \cite{sionna_phy_abstraction}. OFDM's parallel subcarriers capture frequency-selective fading effects through this ESNR metric.

This study focuses on designing a joint agent and semantic encoder-decoder framework to map generated messages into OFDM symbols at the transmitter and recover them at the receiver. The objective is to preserve task performance while reducing bandwidth consumption. Messages transmitted from the BS to the robot are referred to as \textbf{downlink} transmissions, while messages sent from the robot to the BS are referred to as \textbf{uplink} transmissions.

\subsection{Benefits and Challenges of Combining Agents and Semantic Communication}

Integrating AI agents with semantic communication improves information exchange efficiency. Traditional communication schemes transmit generated messages directly, which can be inefficient and semantically redundant. Semantic communication enables agents to extract contextual and task critical information, while agent level processing further refines messages by exploiting background knowledge and omitting irrelevant content.

However, several challenges remain. Semantic encoder-decoders typically operate through end to end training, resulting in implicit representations that are difficult to control and may negatively impact task performance. Moreover, although agents can explicitly process semantics, different agents may interpret the same information inconsistently, leading to misunderstandings. For instance, translation agents may fail to capture cultural nuances. Without standardized semantic representations, interoperability across agents and tasks remains limited, constraining the overall effectiveness of agent semantic communication.

\section{Proposed Framework for Embodied AI Agents}
\label{s3}

This section presents the communication framework between the BS and the robot, which involves multi round interactions between AI agents. To improve transmission efficiency, messages generated by the BS are simplified using the proposed compressor in cooperation with an LLM. The communication process is further designed to adapt to different system requirements. Finally, transmission bandwidth between the client and the server is reduced through a semantic encoder-decoder for specific tasks.

\begin{figure*}
	\centering
\subfigure[]{\includegraphics[width=0.99\linewidth]{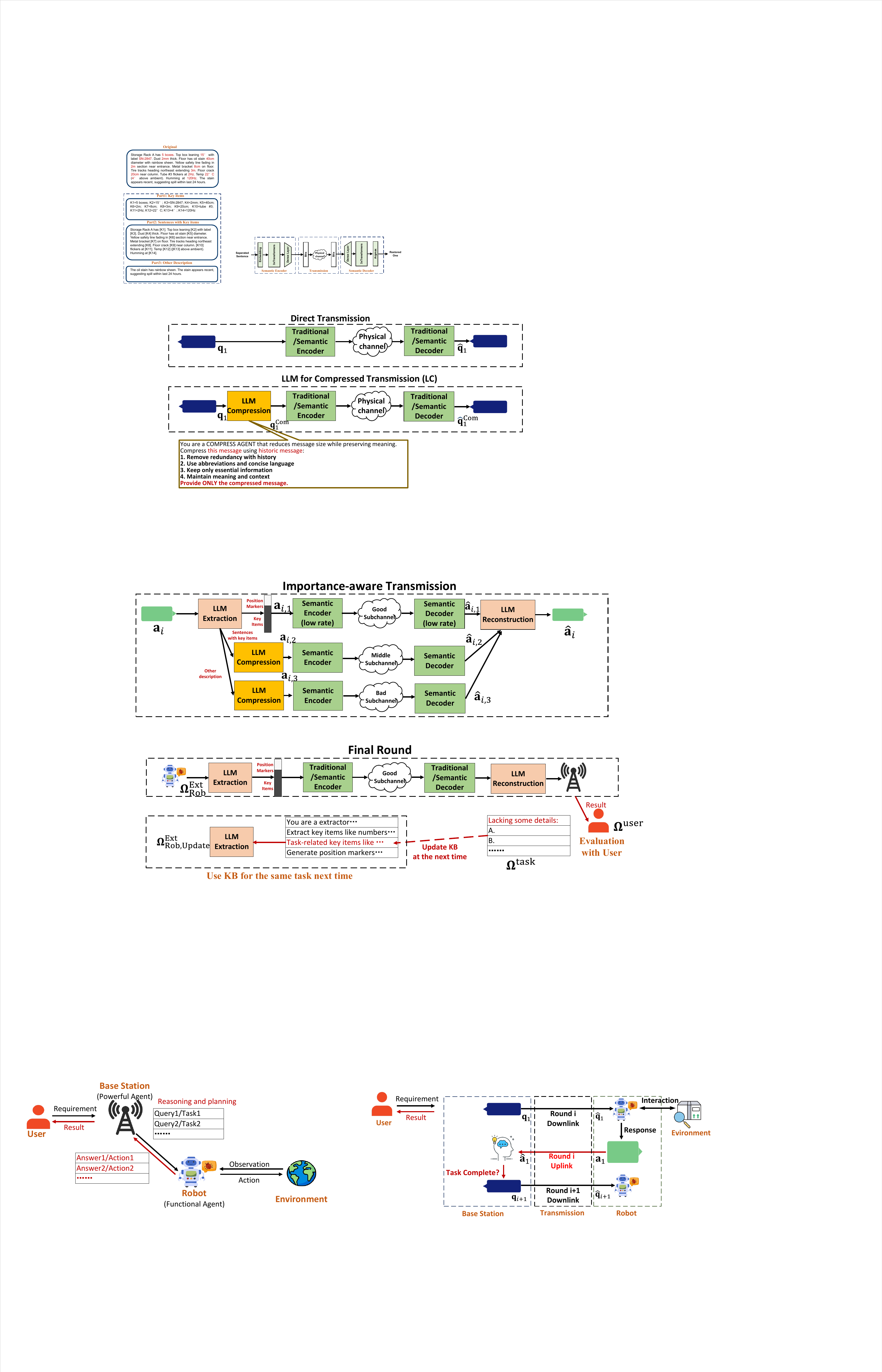}}
\subfigure[]{\includegraphics[width=0.99\linewidth]{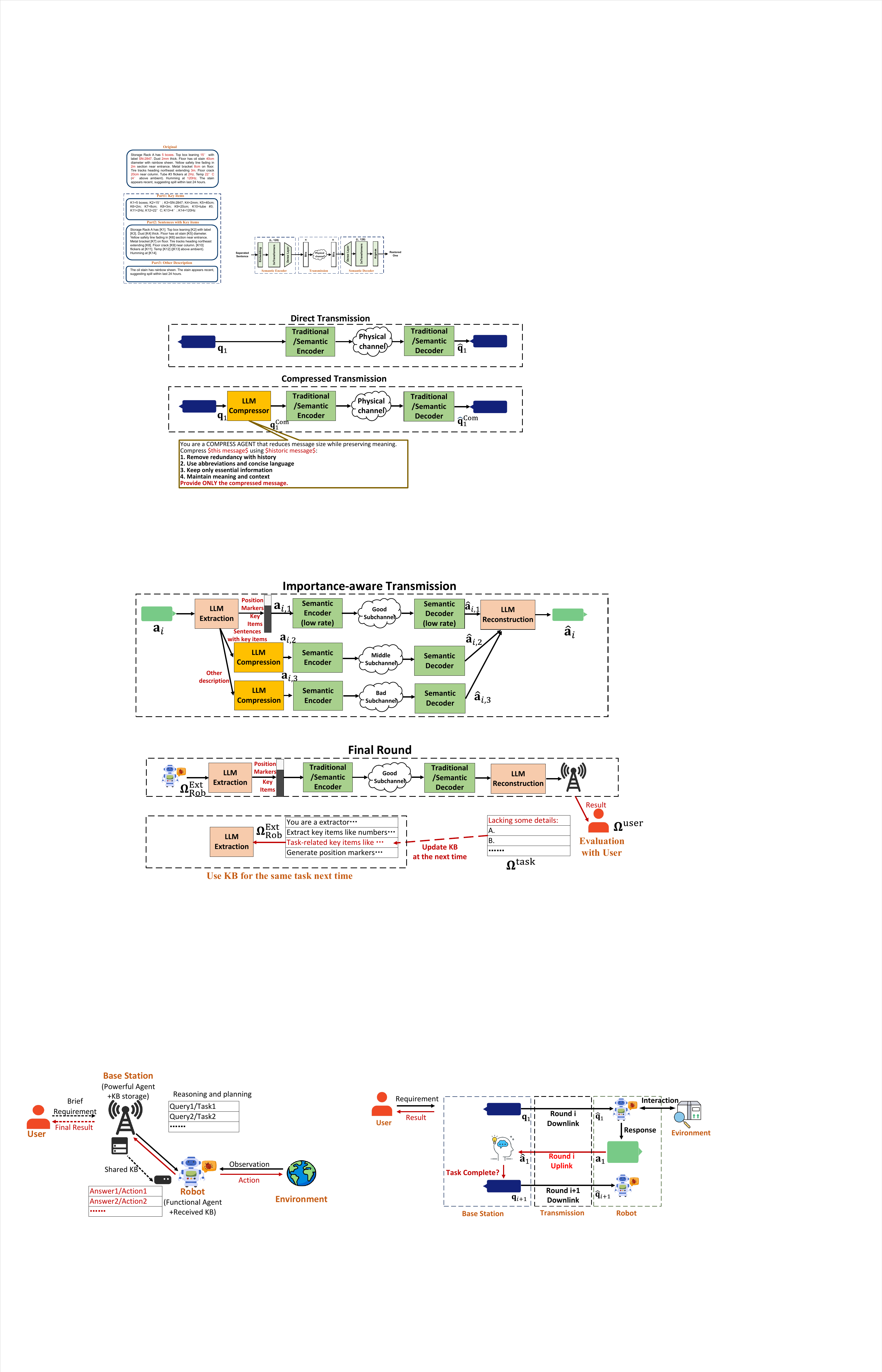}}
	\caption{(a) The proposed LLM based compressor (LC) applied before message transmission with the corresponding prompt. (b) Architecture of the semantic encoder and decoder for text messages.} 
	\label{Transarchi}
\end{figure*}

\subsection{Basic Framework}

 In the context of Fig.~\ref{Overall}, we first define the roles of the BS and the robot, as well as the messaging primitives exchanged between their corresponding agents. The objective is to enable efficient embodied task execution with minimal user involvement. The \textbf{user} provides a concise task requirement to the BS and waits for the final result.

The \textbf{BS} hosts the full LLM and complex agents, which convert the user requirement into executable task steps. To support this process, the BS maintains a powerful KB that stores historical information to learn user preferences and potential environments.

The \textbf{robot} acts as a constrained executor and a lightweight preprocessor. It interacts with the environment through onboard sensors and performs simple instructions based on prompts received from the BS. The robot extracts and organizes task relevant information to generate a preprocessed response. This message is then converted into bits and transmitted to the BS through the uplink wireless channel.

The overall communication process is expressed as
\begin{equation}
    \mathbf{q}_1=\Lambda_{\rm BS}(\mathbf{q}_{\rm U},\mathbf{\Omega}_{\rm BS}),
\end{equation}
where $\mathbf{q}_{\rm U}$ denotes the user requirement, $\mathbf{\Omega}_{\rm BS}$ represents the knowledge stored at the BS, and $\Lambda_{\rm BS}(\cdot)$ is the BS agent that divides the task into multiple steps $\mathbf{q}_i$. 

In the first downlink transmission, $\mathbf{q}_1$ is encoded into OFDM symbols and received by the robot with the ESNR as in  (\ref{ESNR}). The decoded message $\hat{\mathbf{q}}_1$ is used as the robot prompt to interact with the environment. The robot response is generated based on sensor information $\mathbf{\Omega}_{\rm Rob}$ as
\begin{equation}
    \mathbf{a}_1=\Lambda_{\rm Rob}(\hat{\mathbf{q}}_{1},\mathbf{\Omega}_{\rm Rob}), \label{eq2}
\end{equation}
where $\Lambda_{\rm Rob}(\cdot)$ denotes the robot agent. The uplink transmission follows a similar process, yielding the received response $\hat{\mathbf{a}}_1$. The BS agent then determines whether additional steps are required. For the next step, the query is generated as
\begin{equation} 
    \mathbf{q}_2=\Lambda_{\rm BS}([\mathbf{q}_{\rm U},\mathbf{q}_1,\hat{\mathbf{a}}_1],\mathbf{\Omega}_{\rm BS}),
\end{equation}
where $\mathbf{q}_2$ excludes termination indicators such as ``task complete''. The total number of communication rounds is denoted by $R$.

The goal of this framework is to enable reliable and efficient message exchange. By coordinating BS and robot agents, complex tasks can be completed without intermediate user participation. However, LLM based agents may generate outputs with uncontrolled length and redundant information. In addition, wireless channel impairments reduce message reliability and affect task performance. The tradeoff between bandwidth consumption and task accuracy is examined in the following subsections.

\begin{figure*}
	\centering
{\includegraphics[width=0.99\linewidth]{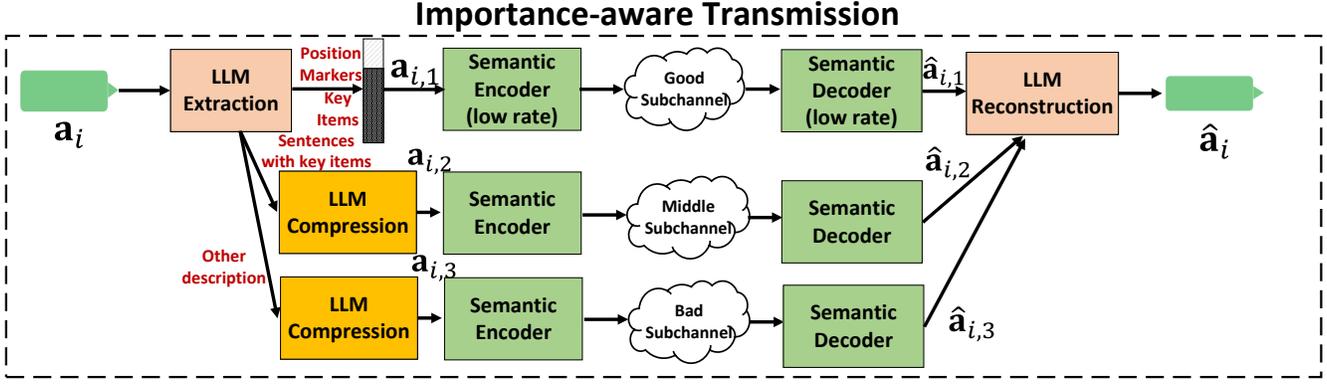}}
	\caption{Illustration of LLM based extraction and importance-aware transmission, where key items with position markers are assigned stronger protection.}
	\label{Imtrans}
\end{figure*}

\subsection{LLM Agent for Message Compression and Semantic Encoder-Decoder}

A direct approach to reducing transmission overhead is to compress text messages using an LLM based agent. As shown in Fig.~\ref{Transarchi}(a), the output message generated by the BS agent, such as $\mathbf{q}_1$, is compressed as
\begin{equation}
    \mathbf{q}_1^{\rm Com}=\Lambda^{\rm Com}_{\rm BS}(\mathbf{q}_1,\mathbf{\Omega}^{\rm Com}_{\rm BS}),
\end{equation}
where only the prompt $\mathbf{\Omega}^{\rm Com}_{\rm BS}$ at the BS needs to be designed. The compressed message is then encoded using either a conventional codec or a semantic encoder-decoder. The received message $\hat{\mathbf{q}}^{\rm Com}_1$ is used as the robot input, as in  (\ref{eq2}).

When a semantic encoder-decoder is selected, a semantic encoder $SC_{\rm en}(\cdot)$ converts $\mathbf{q}_1^{\rm Com}$ into transmitted OFDM symbols, as illustrated in Fig.~\ref{Transarchi}(b). The text is first segmented into sentences. Sentences longer than $L$ words are divided, while shorter sentences are padded with zeros to achieve a fixed length. Each word is mapped to an index using the embedding dictionary. This preprocessing ensures a fixed input size for the semantic encoder. Subsequently, three Transformer blocks are applied with the hidden dimension is 128, followed by a dense layer to reduce the output dimensionality into $n$. The dense layer output is normalized to the range $[0,1]$ using a Sigmoid activation function, and a one bit quantization operation is applied to generate $n$ binary symbols. The decoder is an inverse version of the encoder,  where an argmax function is used for output of the last Transformer block  and choose the reconstructed word from the dictionary.

In this study, the compressor prompt at the BS is illustrated in Fig.~\ref{Transarchi}(a), where historical messages can be utilized. For example, the historical messages up to the $(r-1)$th round are represented as $[\mathbf{q}_1,\ldots,\mathbf{q}_{r-1}; \hat{\mathbf{a}}_1,\ldots,\hat{\mathbf{a}}_{r-1}]$. In contrast, the compressor at the robot does not store historical messages and compresses content using only the inherent knowledge of the LLM. This design reduces computational burden at the robot and helps protect user historical information.

Within this compression and transmission framework, information loss may occur at three stages:
\begin{enumerate}
    \item \textbf{LLM compressor:} Unimportant content may be removed during length reduction, with detailed descriptions typically discarded first.
    \item \textbf{Semantic encoder-decoder:} As a trainable codec, reconstruction errors are inevitable. Fixed bit length  constraints may further reduce accuracy.
    \item \textbf{Wireless channel:} Channel degradation reduces transmission reliability. Common mitigation approaches include lower code rates for conventional channel coding or longer bit sequences for semantic coding.
\end{enumerate}

Due to the distinct characteristics of these stages, increasing bandwidth is often considered a direct method to alleviate information loss. However, the transmitted content can instead be partitioned and processed according to task requirements. This flexible strategy is discussed in the following subsections.

\begin{figure}
	\centering
{\includegraphics[width=0.9\linewidth]{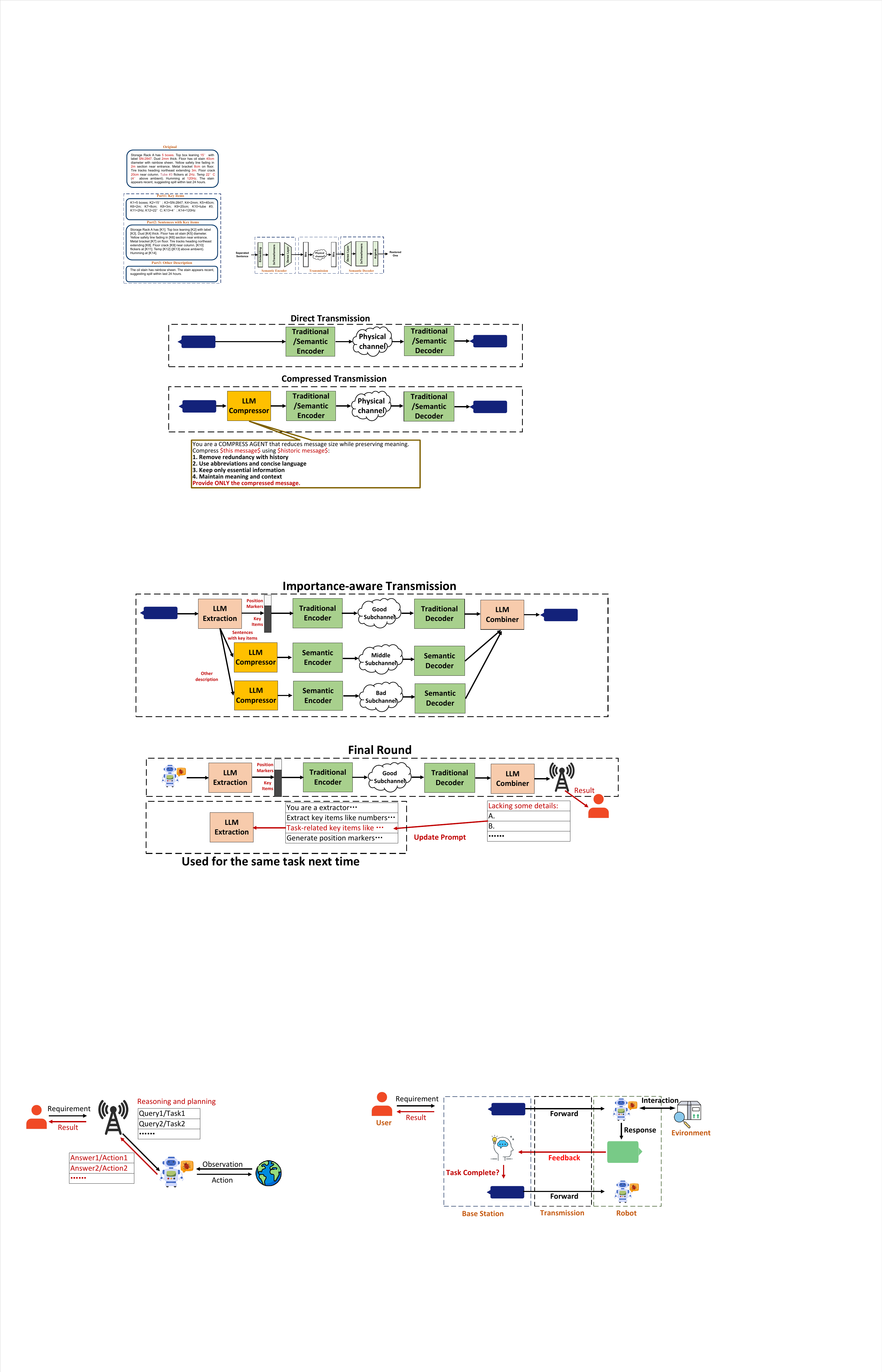}}
	\caption{The message sent from the robot to the BS at the round 1, which responses to the surrounding inspection query of the BS.}
	\label{Imtrans_ex}
\end{figure}
\begin{figure*}
	\centering
	{\includegraphics[width=0.99\linewidth]{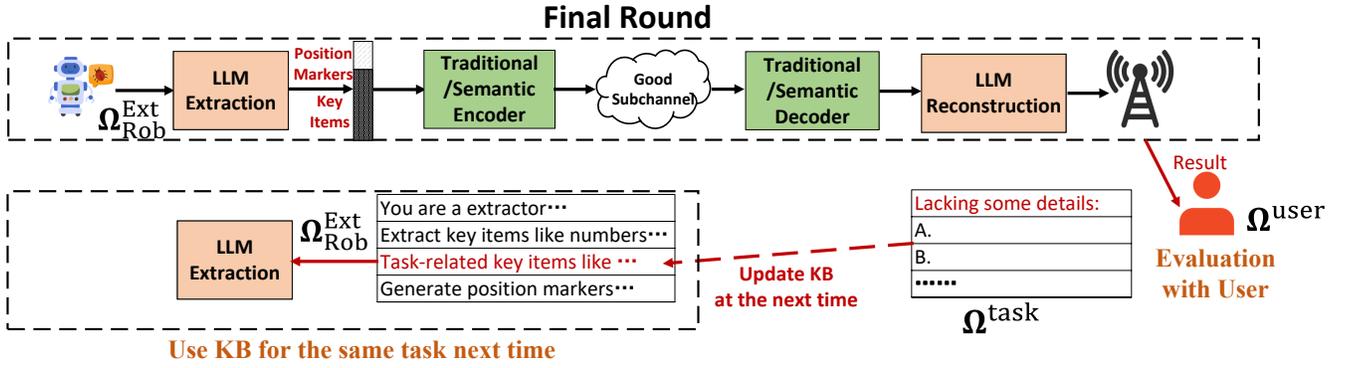}}
	\caption{The framework of the KB update process based on the user's evaluation of the last same task. }
	\label{KBU}
\end{figure*}

\subsection{LLM Agent for Message Extraction and Importance-aware Transmission}
Instead of directly increasing bandwidth to mitigate task performance degradation, importance-aware transmission adapts transmission strategies according to content importance and channel conditions. This approach provides stronger protection for critical information and helps preserve task performance. As shown in Fig.~\ref{Imtrans}, three parallel transmission processes are established, where subcarriers are grouped into three subchannels based on their signal to noise ratios and the lengths of the corresponding message components. More important information is transmitted over higher quality subchannels. This design targets the frequency selective channel considered in this study. Alternative protection mechanisms, such as lower order modulation or reduced code rates, may also be applied to safeguard critical content.

Before transmission, an LLM based extractor divides each message into three parts: key items, sentences containing key items, and remaining sentences. An example workflow is provided below. The first part of the $i$th round message $\mathbf{a}_{i}$ transmitted from the robot to the BS is expressed as
\begin{equation}
\mathbf{a}_{i,1}= \left[\Lambda^{\rm Ext}_{\rm Rob}(\mathbf{a}_i,\mathbf{\Omega}^{\rm Ext}_{\rm Rob}),\mathbf{POS}_{\mathbf{a}_i} \right],
\end{equation}
where $\mathbf{a}_{i,1}$ consists of key items and their corresponding position markers $\mathbf{POS}_{\mathbf{a}_i}$, and $\Lambda^{\rm Ext}(\cdot)$ denotes the LLM based extraction function. The second part is defined as
\begin{equation}
    \mathbf{\bm a}_{i, 2}=\Lambda^{\rm Com} {\left(\Lambda^{\rm Ext}(\mathbf{\bm a}_i,\mathbf{\bm a}_{i, 1},\mathbf{POS}_{\mathbf{\bm a}_i}]),\mathbf{\Omega}^{\rm Com}_{\rm Rob} \right)},
\end{equation}
where sentences containing key items are extracted and the key items are replaced by their position markers. The LLM based compressor at the robot further reduces this content to generate $\mathbf{a}_{i,2}$. The remaining sentences are collected and compressed to form the third part $\mathbf{a}_{i,3}$. Fig.~\ref{Imtrans_ex} illustrates this extraction process using the first round message from the robot to the BS as an example. Fourteen key items are marked in red and their corresponding position markers [K1]-[K14] are inserted in the Part1 and the Part2.

The three message parts are transmitted through different transceivers. For the first part, a semantic encoder-decoder with $n'$ per sentence is applied to ensure reliable transmission, while favorable subchannels provide additional protection. The remaining parts are transmitted using LLM compression combined with a semantic encoder-decoder with $n$ bits per sentence. In order to protect the key items better, the low code rate should be used, i.e., $n'>n$.

At the receiver, an LLM based reconstruction process restores the complete message. First, key items are inserted into their original positions using the position markers. Then, sentences containing key items are merged with the remaining content to generate the reconstructed message
\begin{equation}
    \hat{\mathbf{a}}_{i}=\Lambda^{\rm Rec}{\left([\hat{\mathbf{a}}_{i,1},\hat{\mathbf{a}}_{i,2},\hat{\mathbf{a}}_{i,3}]\right)},
\end{equation}
where $\hat{\mathbf{a}}_{i,1}$ denotes the received first part from the semantic decoder. This reconstruction process relies on the semantic understanding capability of the LLM and does not require strict alignment.

With importance-aware transmission, task critical information typically occupies only a small fraction of the message, resulting in a modest increase in bandwidth to accommodate position markers. However, due to the reliance on general LLM knowledge, detailed descriptions may be partially omitted with the wrong choices of the key items. While effective in most scenarios, this approach may lead to the loss of fine grained details in certain task specific cases.

\subsection{Update Knowledge for Specific Tasks}

Given the strong computational capability of the BS, maintaining a KB is an effective approach to further improve task performance based on historical executions. The KB should be efficiently established when encountering a new task or a new user.

The BS receives task requirements from the user and checks whether the corresponding user profile and task knowledge are already stored. If such knowledge is unavailable, the user is required to perform a detailed evaluation of the execution results and provide feedback. As shown in Fig.~\ref{KBU}, after the final round of a previously unseen task, the overall task performance is assessed. The figure illustrates an example of the LLM based extraction module at the robot, where missing details are identified by comparing the user requirements with the robot responses and subsequently stored at the BS.

When the same task is requested again, the stored details are transmitted in advance to guide the robot extractor. This process is autonomously initiated by an LLM based evaluator, which analyzes missing information by aggregating historical messages at the BS together with the user feedback $\mathbf{\Omega}^{\rm user}$, yielding 
\begin{equation}
    \mathbf{\Omega}^{\rm task}=\Lambda^{\rm Eva}{\left([\hat{\mathbf{a}}_1,\ldots,\hat{\mathbf{a}}_R; \mathbf{q}_1,\ldots,\mathbf{q}_R],\mathbf{\Omega}^{\rm user} \right)},
\end{equation}
where $\mathbf{\Omega}^{\rm task}$ represents task specific details with representative examples. This information is then incorporated into the robot knowledge for future executions as

\begin{equation} 
    \mathbf{\Omega}^{\rm Ext}_{\rm Rob} \leftarrow [\mathbf{\Omega}^{\rm Ext}_{\rm Rob}, \mathbf{\Omega}^{\rm task}].
\end{equation}

\subsection{Semantic Coding versus LLM-based Methods}
Unlike existing semantic communication approaches, LLM-based compression and extraction are introduced as agent level mechanisms to reduce transmission overhead. Although both semantic encoders and LLM-based compressors aim to reduce text while preserving meaning, they differ in objectives, design principles, and outputs, as summarized below.

\begin{enumerate}
    \item \textbf{Semantic encoder:} A semantic encoder typically maps inputs into implicit and model based representations, such as dense vectors or key value features, which are optimized for downstream reconstruction. It emphasizes semantic fidelity and commonly relies on end to end training. However, the fixed network architecture and output dimensionality limit its flexibility.

    \item \textbf{LLM-based method:} An LLM-based method generates human readable and lossy text representations, such as summaries, key points, or rationales, guided by prompts and prior knowledge. It trades comprehensive coverage for task specific effectiveness.
\end{enumerate}

Semantic coding offers low latency retrieval and deterministic structure but does not explicitly reduce text length. In contrast, LLM-based methods provide adaptable abstraction, task-aware expression, and cross task generalization. When combined, semantic encoders can benefit from LLM-based processing, as LLMs can generate shorter and more informative text through reasoning and planning.

\section{Simulation Results}
\label{s5}
This section discusses the simulation results of the proposed  transmission workflows. After introducing the testing cases and environment settings, we first evaluate the performance and the bandwidth cost of the basic communication with the proposed LLM compression. Then, we demonstrate the effectiveness of the proposed importance-aware transmission with LLM extraction and reconstruction methods. Meanwhile, the performance of the proposed methods with knowledge update can further benefit the entire bandwidth cost with only a little performance loss.  Finally, different usages of LLM-based and semantic methods are discussed.

\subsection{Two Different Cases and Environment Settings}

This study focus exclusively on evaluating wireless  communication between agents in specific embodied settings, rather than pursuing broad generalization across tasks. Consequently, we just generate two simple scenarios from the existing datasets\cite{das2018embodied,shridhar2020alfred} : 1) a warehouse robot inspection scene, and 2) a household cleaning scene. These environments are good cases to evaluate the transmission accuracy effect between agents with several communication rounds. The agents at the BS are established based one Claude-Sonnet-4.5 while the those at the robot are based on GPT-4o.
\begin{enumerate}
    \item \textbf{Case1} usually requires  \textbf{long communication messages}. Inspection task requires the robot to  detect the current conditions using their sensors. Thus, the response may be long with many detailed descriptions. 

 \textbf{Simulated Environment:} The scenario depicts a warehouse inspection conducted by a mobile robot  equipped with camera, thermal, LiDAR, and audio sensors. Within a 12m × 15m area, the storage racks are arranged in two columns that form aisles. The robot patrols to identify safety and maintenance issues. Some findings are randomly generated by an environment agent. 

 \textbf{Result:} We randomly generate around 20 sensor numbers and several conditions, such as oil stain and burned-out lightning for testing.

 \item \textbf{Case2} usually requires \textbf{short message} because a household cleaning robot only reports its execution status of the command.

 \textbf{Environment:} The household cleaning robot task involves a living room cleanup in a space. The robot must relocate some items. The task follows three phases: perception, sequential pickup and final verification. 

 \textbf{Result:} We randomly generate around 20 items at different coordinates for testing.

\end{enumerate}

For comparison, the conventional baseline consists of Huffman coding combined with LDPC channel coding with a code rate of 1/2. The 4-QAM is used for the OFDM transmission, where the LDPC code can guarantee nearly no bit errors when SNR is 10 dB. The settings of the semantic communication for text is similar to \cite{xie2020deep,jiang2022deep}. The SC with $n=1000$ bits for $L=30$ words is applied, where the word error rate is smaller than 1\% when SNR is larger than 5 dB and the word error rate reaches about 10\% when ${\rm SNR} = 0$~dB. The SC for key items using $n'=2000$ bits for 30 words and can always maintain the word error rate is smaller than 1\% when ${\rm SNR} > 0$~dB.

\begin{figure*}[h]
	\centering
	\subfigure[Case1]{		
		{\includegraphics[width=0.49\linewidth]{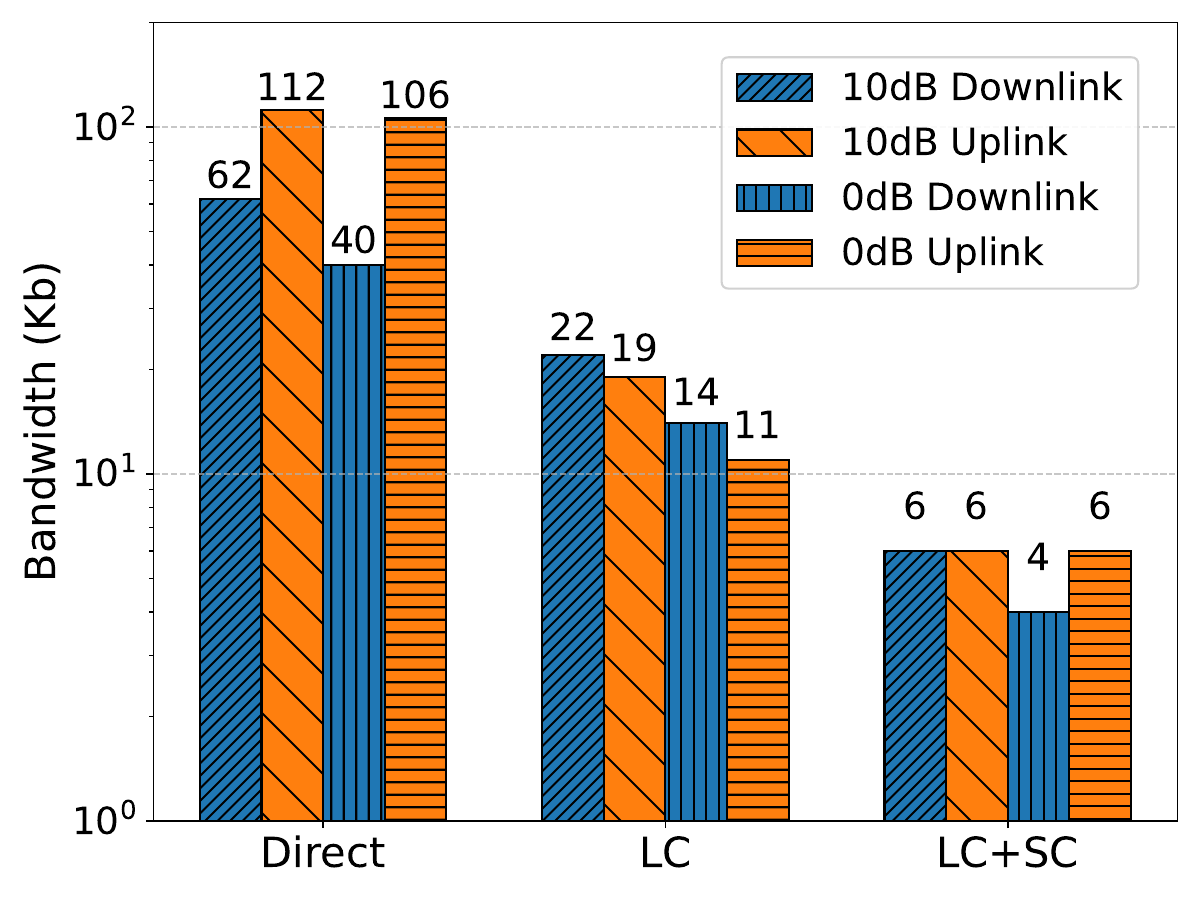}}}
	\subfigure[Case2]{
		{\includegraphics[width=0.49\linewidth]{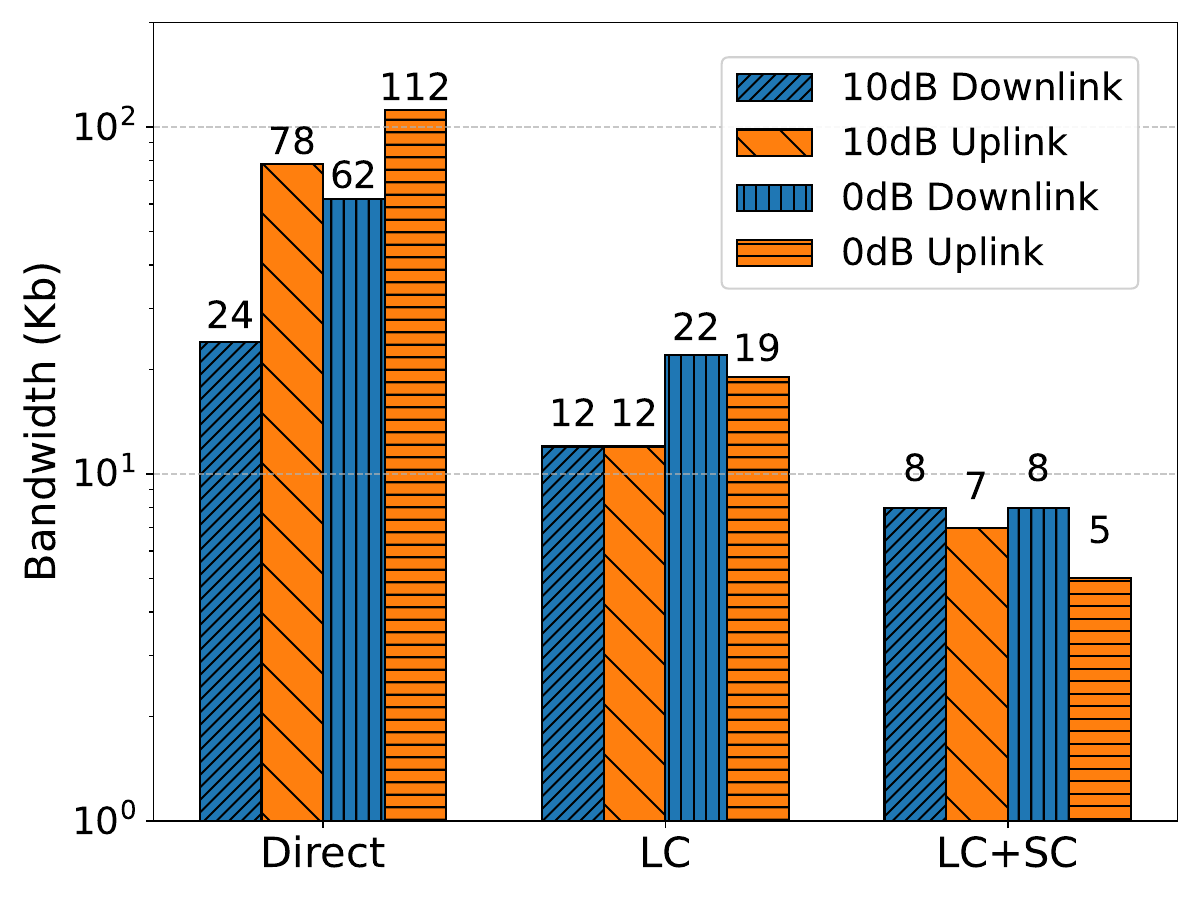}}}
	\caption{Bandwidth cost of different scenarios only using LLM compression.}
	\label{Secasce}
\end{figure*}

Different from  the conventional metric, two modern metrics are introduced to evaluate performance across different tasks:  
\begin{itemize} 
	\item \textbf{Successful Rate (SR):} SR in WebArena and AgentBench evaluates an LLM-as-Agent’s ability to complete end-to-end tasks by measuring  whether outputs achieve task-specific goals and include all critical information\cite{zhou2023webarena}.  In the following test, the number of critical goals are calculated according to the specific scenario.

	\item \textbf{Distinct-1:} A quantitative metric widely used in natural language processing to evaluate the diversity of text generation models by calculating the ratio of unique n-grams to total n-grams in generated outputs, with higher values indicating greater diversity and lower redundancy \cite{li2016diversity}. 
\end{itemize}

\subsection{Performance of the LLM Compression}
To evaluate the communication performance of the proposed framework, we compare different methods in two cases under different channel conditions. The direct communication using conventional Huffman+LDPC coding  is called “Direct”. The competing method using LLM compression and conventional coding is called “LC” while that using the semantic encoder-decoder is called “LC+SC”. In order to avoid the uncontrollable rounds due to the large noise, the maximum number of communication rounds is set to 5.

For the inspection task, Fig.~\ref{Secasce}(a) presents different phenomena. In order to  collect the necessary information from the scene, the BS and the robot need more words to describe themselves, where the total bandwidth cost is larger than the previous task. Owing to this characteristic, the LLM compression can save more transmission resources. The LC requires around 1/3 downlink and 1/8 uplink resources compared to the direct one. However, such prominent bandwidth benefits bring potential performance risks. Meanwhile, the bandwidth costs of direct method under 0 and 10 dB SNRs are similar because both tests reach our maximum 5 rounds limitation and the BS cannot ask the robot to re-collect the previous information. The following task performance will further demonstrate the effects of different transmission strategies.

As shown in Fig.~\ref{Secasce}(b), the LC can reduce most of the transmission overhead while the LC+SC can further use the fewest resources.   When ${\rm SNR} = 10$~dB, the LC uses only 1/2 downlink and 1/6 uplink bandwidth compared to the direct one.  In the direct communication, the robot tends to transmit longer message than the BS because the BS knows the whole task and reaches its goal step-by-step. In contrast, the robot always tries to report all its information, which is considered to reach a better task performance. Moreover, the errors in the message also leads more rounds of communication or longer message in the next round.  When ${\rm SNR} = 0$~dB, the bandwidth cost of direct one becomes larger and the LLM compression has the space to save more bandwidth. The increasing bandwidth of the LC is less than the direct one. The LC+SC under 0 dB SNR has similar bandwidth cost as that under 10 dB SNR due to the advantage of semantic encoders in high-noise conditions.

\begin{table}
    \centering    
    \caption{Task performance of \textbf{Case1}.}
    \begin{tabular}{crrcc}
        \toprule
        Method & SNR & SR$\uparrow$ & Distinct-1$\uparrow$ & Complete \\ \midrule
        \multirow{4}{*}{Direct} 
            & 10 dB & \textbf{100} & 0.342 & yes \\ \cmidrule{2-5}
            & 5 dB & 0 & / & no \\ \cmidrule{2-5}
            & 0 dB & 0 & / & no \\ \midrule
        \multirow{4}{*}{LC} 
            & 10 dB & 95 & 0.517 & yes \\ \cmidrule{2-5}
            & 5 dB & / & / & no \\ \cmidrule{2-5}
            & 0 dB & / & / & no\\ \midrule
        \multirow{4}{*}{LC+SC} 
            & 10 dB & 94 & \textbf{0.672} & yes \\ \cmidrule{2-5}
            & 5 dB & 85 & \textbf{0.451}& yes \\ \cmidrule{2-5}
            & 0 dB & \textbf{75} &\textbf{ 0.324} & no \\ 
        \bottomrule
    \end{tabular}
    \label{Metric2}
\end{table}

The inspection task results in TABLE \ref{Metric2} express the drawbacks of the simple LC with losing details. The direct method can only complete the task when SNR=10 dB, achieving 100\% SR among all competing methods. When SNR decreases, the direct method cannot complete the task. That means the transmission loss prevents communication between the BS and the robot at this condition. The LC+SC is regraded by the BS as having completed the task when SNR is 10 and 5 dB. However, the final evaluation finds that there are around some details lost.  This phenomenon demonstrates that this case with rich details is difficult for compression and transmission. The direct method shows similar distinc-1 values at different SNRs. In contrast, the LC-based methods lose their diversity with increasing noise because the wrong information disturbs reliable communication.

\begin{table}
    \centering    
    \caption{Task performance of \textbf{Case2}.}
    \begin{tabular}{crrcc}
        \toprule
        Method & SNR & SR$\uparrow$ & Distinct-1$\uparrow$ & Round$\downarrow$ \\ \midrule
        \multirow{4}{*}{Direct} 
            & 10 dB & 100 & 0.151 & \textbf{2} \\ \cmidrule{2-5}
            & 5 dB & 0 & / & 5\\ \cmidrule{2-5}
            & 0 dB & 0 & / & 5 \\ \midrule
        \multirow{4}{*}{LC} 
            & 10 dB & 100 & \textbf{0.562 }& \textbf{2} \\ \cmidrule{2-5}
            & 5 dB & 0 & / & 5 \\ \cmidrule{2-5}
            & 0 dB & 0 & / & 5\\ \midrule
        \multirow{4}{*}{LC+SC} 
            & 10 dB & 100 & 0.421 & 3 \\ \cmidrule{2-5}
            & 5 dB & 100 & \textbf{0.351}& \textbf{4 }\\ \cmidrule{2-5}
            & 0 dB & \textbf{87} & \textbf{0.488} &\textbf{3} \\ 
        \bottomrule
    \end{tabular}
    \label{Metric1}
\end{table}

TABLE \ref{Metric1} shows different  results when completing the cleaning task with the LLM compression. In this case, the SR performance is not affected by the reduced message because the instruction of the BS is clear and the robot only reports its action one-by-one. All the objects randomly generated by the simulated environment are reached. Meanwhile, the diversity of the message (distinc-1) is increased, which means the redundancy of the transmitted sentences is low and the repeated words are few. When SNR goes lower, the number of transmission rounds increases because the BS and the robot may receive wrong information and need to check again, where the distinc-1 also decreases.

\begin{figure}
	\centering
	
		\includegraphics[width=0.99\linewidth]{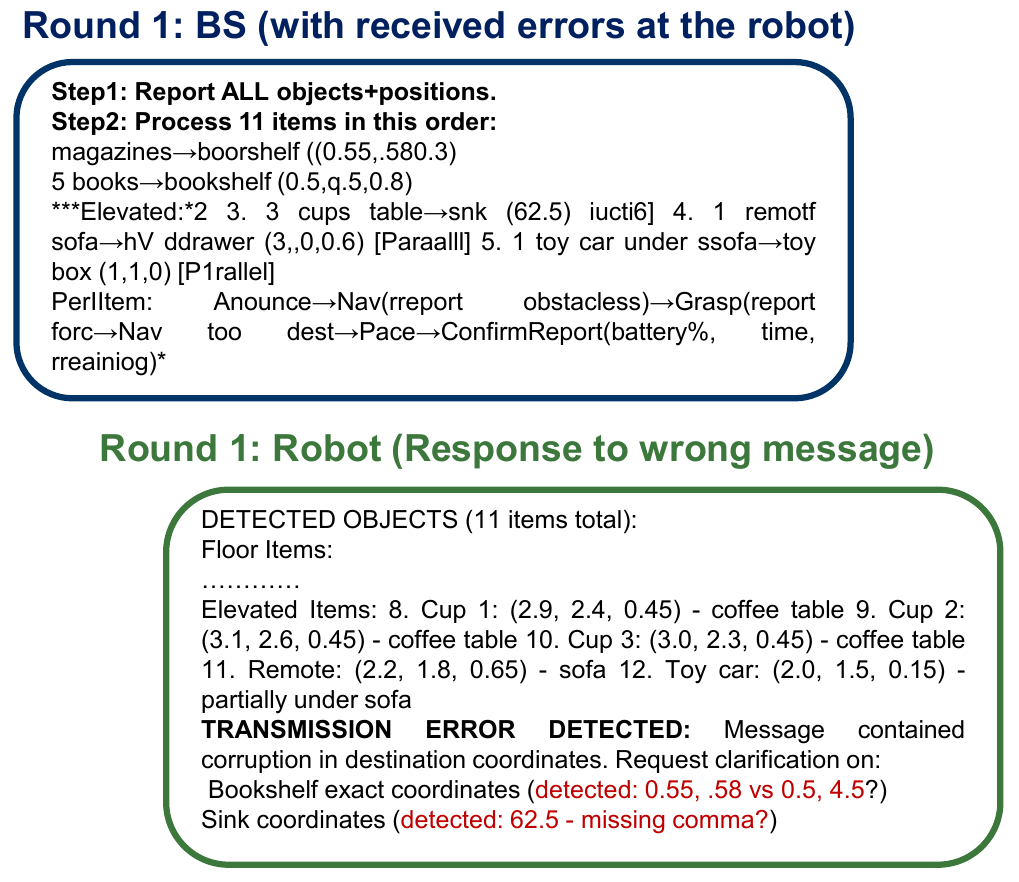}

	\caption{An example of the messages sent between the BS and the robot agents in Case2.}
	\label{EX_A}
\end{figure}

Fig.~\ref{EX_A} provides a part of the communication examples in \textbf{Case2}. In the  round 1, the BS provides some instructions with some objects' positions. The robot executes accordingly and  then reports the current conditions. Due to the transmission noise, some characters in the BS are changed and bring wrong meaning to the robot. However, the robot learns the surrounding environment and detects these errors, such as wrong positions. The robot asks the BS to confirm the wrong information and thus the more rounds will be started.

In summary, the introduction of the LC provides an effective way to reduce bandwidth in an explicit mode, where the transmitted message is shorten to preserve the core information. Meanwhile, the application of SC improves the task performance when the SNR is low. However, this method still brings some information loss and some errors leading to misunderstanding. The LC using the default prompt is suitable for some instructions as in Case2 while Case1 requires more details for the next step. In the following subsections, the improvement strategy  for specific detail transmission is discussed.

\subsection{Performance of  Importance-aware Transmission and KB Update}
This subsection further investigates the enhanced solution for the agent communication with the ``LC+SC'', where all methods use semantic encoder-decoders here. The LC+SC improved by importance-aware transmission is called ``LC+SC(Im)'' while that using the KB update method is called ``LC+SC(Im+KB)''. The Case1 is studied because its performance cannot be well protected in the above tests.

  \begin{figure}
	\centering

		{\includegraphics[width=0.99\linewidth]{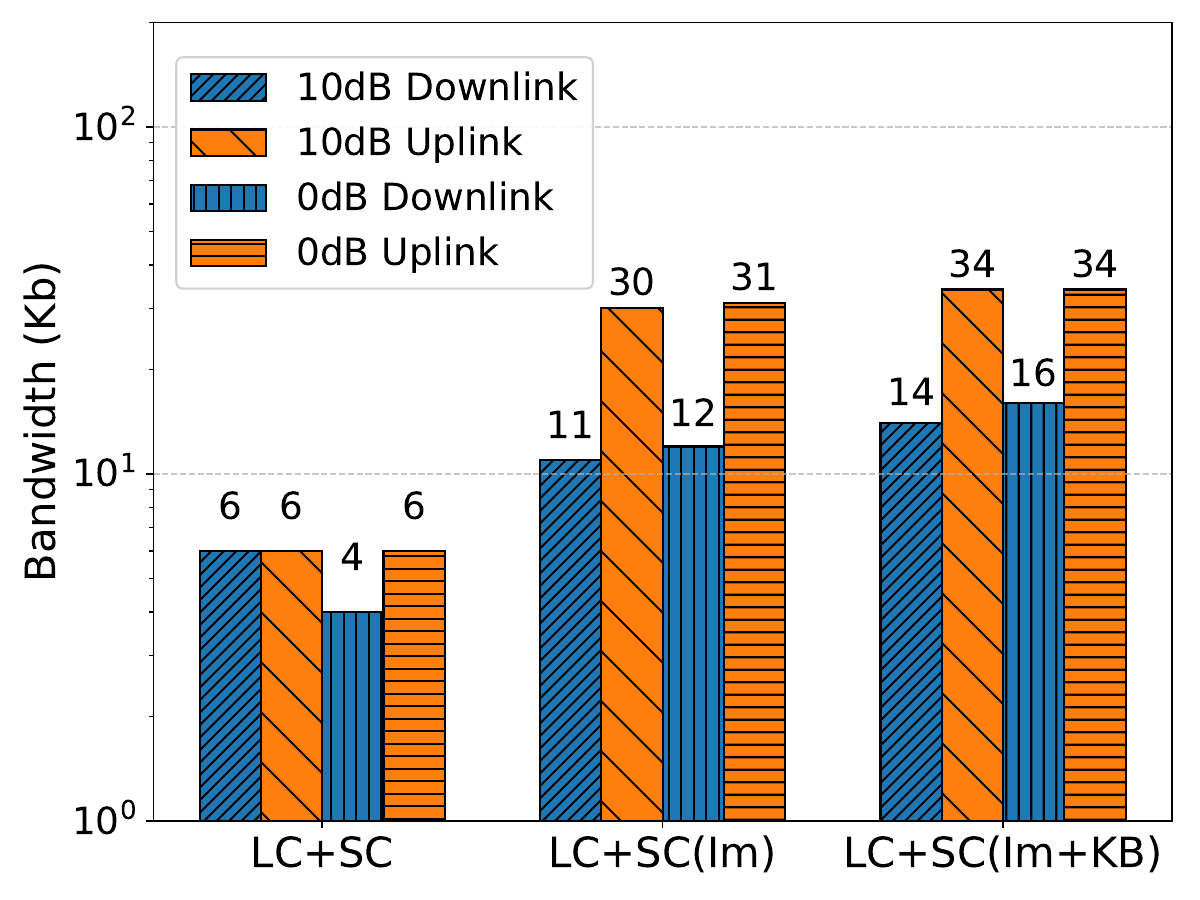}}
\caption{Bandwidth cost of the LC+SC with the improved solutions, where the semantic encoder-decoder is always applied here. }
	\label{SecC_per}
\end{figure}

Fig.~\ref{SecC_per} shows the bandwidth cost of the competing methods, where the LC+SC requires the fewest transmission resources. When the importance-aware method is used in LC+SC(Im), the bandwidth cost increases, but it is still much lower than that of the direct one in Fig.~\ref{Secasce}(b). The downlink transmission from the BS to the robot requires fewer bits than the uplink because the robot's response usually contains more numeric values and specific details. This means the uplink message contains more key items and the LC+SC cannot compress this part. When the KB is inserted into LC+SC(Im+KB), slightly more bits  are transmitted, which implies that several sentences containing the specific meanings are better protected. Besides, the KB should be transmitted at the beginning of the whole transmission process, which will also cost hundreds of bits.

\begin{table}
    \centering    
    \caption{Task performance of Case1 using improved solutions.}
    \begin{tabular}{crrcc}
        \toprule
        Method & SNR & SR$\uparrow$ & Distinct-1$\uparrow$ & Complete \\ \midrule
        \multirow{4}{*}{LC+SC} 
            & 10 dB & 94 & \textbf{0.672} & yes \\ \cmidrule{2-5}
            & 5 dB & 85 & \textbf{0.451} & yes \\ \cmidrule{2-5}
            & 0 dB & 75 & 0.324 & no \\ \midrule
        \multirow{4}{*}{LC+SC(Im)} 
            & 10 dB & 82 & 0.423 & yes \\ \cmidrule{2-5}
            & 5 dB & 82 & 0.443 & yes \\ \cmidrule{2-5}
            & 0 dB & 82 & 0.376 & yes\\ \midrule
        \multirow{4}{*}{LC+SC(Im+KB)} 
            & 10 dB & \textbf{100} & 0.406 & yes \\ \cmidrule{2-5}
            & 5 dB & \textbf{100} & 0.393& yes \\ \cmidrule{2-5}
            & 0 dB &\textbf{ 91} & \textbf{0.389} & yes \\ 
        \bottomrule
    \end{tabular}
    \label{Metric3}
\end{table}

Table~\ref{Metric3} compares the performance of different strategies. The LC+SC still reaches the best distinct-1 score because it transmits the fewest bits. In contrast, it loses a few details and its SR is around 94\%. Then, the SR of the LC+SC goes low with the SNR decreases. Because whether the task is completed is  decided by the BS, this evaluation cannot notice the detail loss and the task completion is acknowledged when SNR is 10 and 5 dB.  For the LC+SC(Im) methods, the BS always judges that the task is completed. However, the lack of task-specific knowledge makes the LC+SC(Im) consistently ignore some part of information and its SR cannot reaches the best. The LC+SC(Im+KB) repairs this problem and its SR can reach 100\% when SNR is high. When SNR is 0 dB, the LC+SC(Im+KB) can still achieve  a good SR and only a few details is blocked by the noise.

\begin{figure}
	\centering

		\subfigure[]{\includegraphics[width=0.99\linewidth]{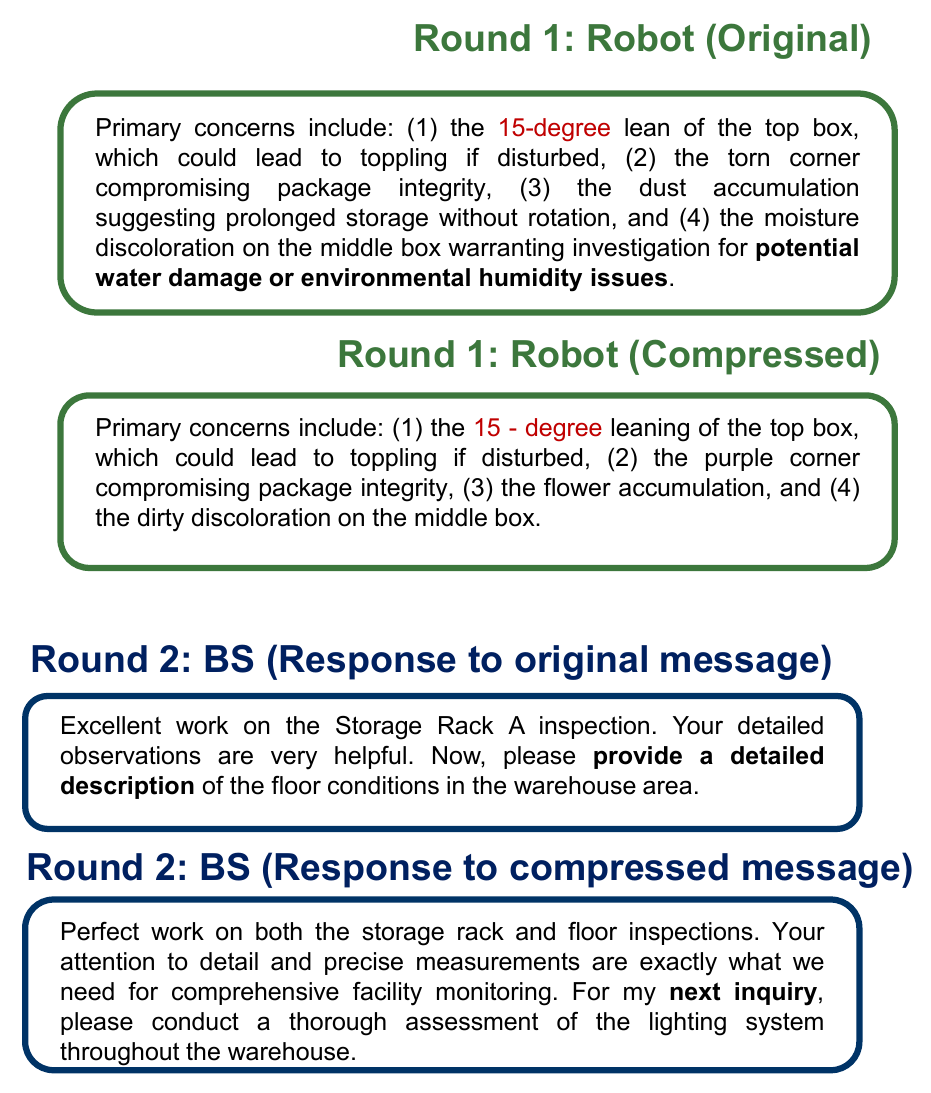}}\\
        \subfigure[]{\includegraphics[width=0.9\linewidth]{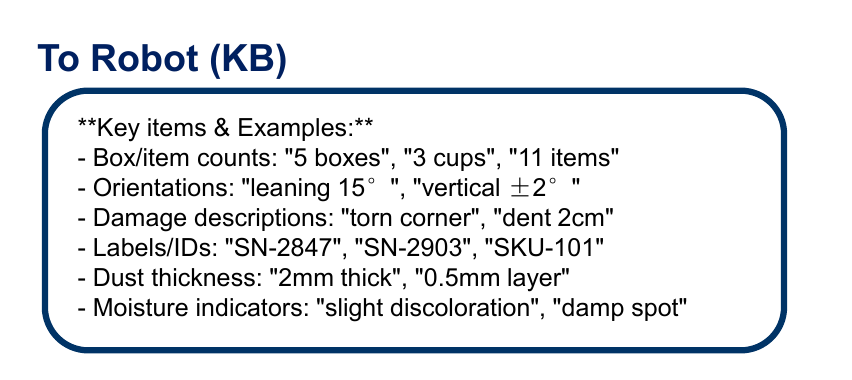}}\\
	\caption{(a) An  Case1 example of the round 1 uplink transmission, where the BS analyzes the message from the robot and then gives the next step. (b) The established KB with the losing details in the compressed message. }
	\label{EX_B}
\end{figure}

The examples in Fig.~\ref{EX_B} further explains the function of the proposed methods.  As shown in Fig. \ref{EX_B}(a), the robot response of the first round contains some detailed description, such as the (2-4) sentences, while the common LLM-based extraction only considers that the specific number  
in the (1) sentence is a key item.  Then, the transmitted message compress the (2-4) and the errors also appear. The BS message in the second round of the direct method usually further investigates the detailed issue based on the robot's report from the previous round. In contrast, receiving the message without some details, the BS may consider that the last inquiry is successfully completed and start the next inquiry. This example shows the reason why some information cannot be known by BS with the application of LC. 
 
 Fig. \ref{EX_B}(b) shows the KB update, where the missing details can be found with the help of the user. The KB to robot records the true important items in this environment.  In the KB, some problems, such as the dust and the moisture, and their  examples are stored in the BS and the LLM-based extraction will not miss these key items in the next time, where the KB will be sent the robot in advance.

In summary, the proposed importance-aware transmission  can protect the key items when the transmission overhead is reduced. That is a suitable solution for the tasks with rich  details, such as inspection. For a specific environment, the required details can be well-preserved with the correct KB.

 \begin{figure}
	\centering
	\subfigure[]{		
		{\includegraphics[width=0.9\linewidth]{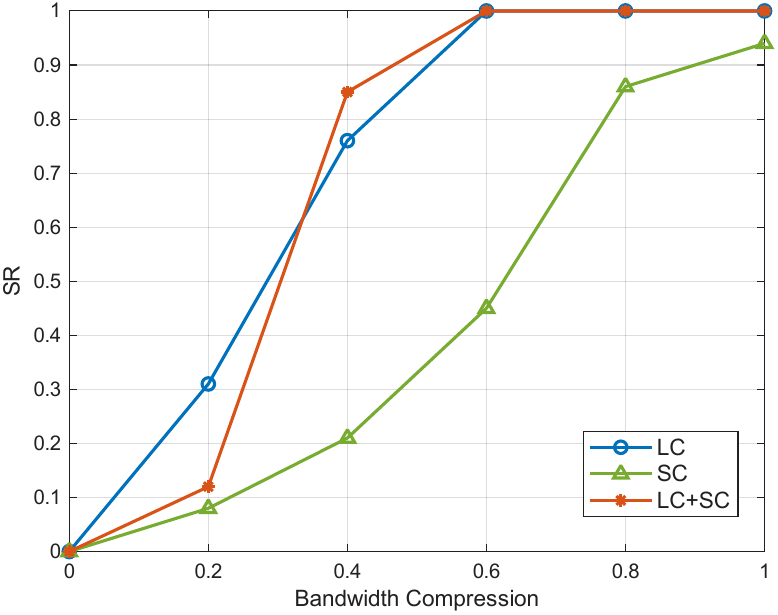}}}\\
	\subfigure[]{
		{\includegraphics[width=0.9\linewidth]{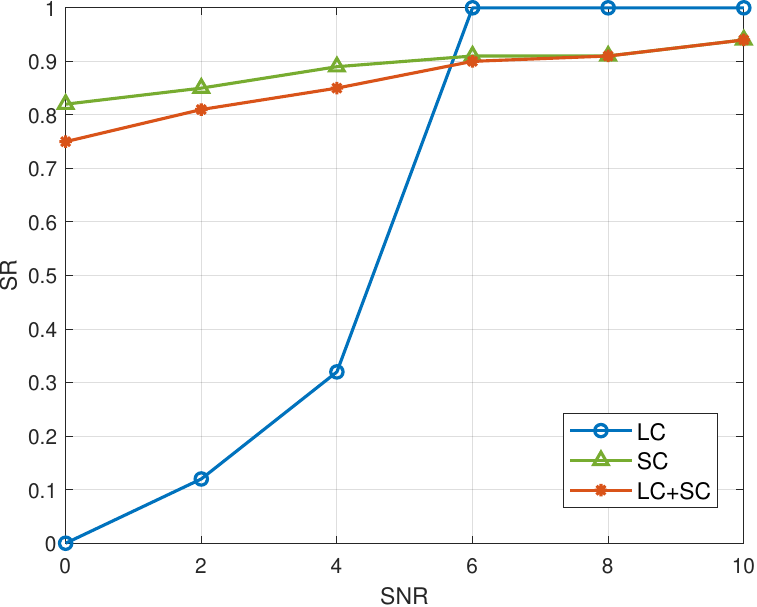}}}
	\caption{Ablation Study for combining LLM and sematic methods for message processing.}
	\label{abastudy}
\end{figure}

\subsection{Ablation Study for LLM compression and Semantic encoder-decoder}

The motivation of combining some LLM methods with the existing semantic encoder-decoder relies on the different advantages of the two directions. The LLM-based one is good at planning and reasoning, which can provide an explicit way to organize the entire message and reduce the communication overhead. The semantic encoder-decoder usually coverts the message into some semantic tokens implicitly.

Fig. \ref{abastudy} shows different characteristics of the LC and SC methods. The bandwidth ratio is calculated from  the original bandwidth and the compressed one. In particular, the original bandwidth of the SC  is about 1000 bits for a sentence, where the word error rate is below 1\%. For the LC method, the original bandwidth is the number of bits of the message encoded by Huffman. For the LC+SC method, the original bandwidth refers to the size of the output message from the LC before it is encoded by the semantic encoder.

As shown in Fig. \ref{abastudy}(a), when the bandwidth ratio is larger than 0.6, the LC does not lose the SR performance and that represents about 40\% content length outputted by the agent can be reduced without key items being lost. Then, the SR of the LC decreases as the bandwidth goes down. As the implicit method, the SC cannot accurately avoid the loss of core information and thus the SR performance always rises with increasing bandwidth. The LC+SC method only uses the LC for compression when the bandwidth ratio is larger than 0.6 and then reduces the bit length of the SC when the lower bandwidth ratio is required.  This method reaches the best SR when the bandwidth ratio is larger than 0.3. In fact, considering the superiority of SC in reducing transmission overhead, the LC+SC methods always achieve the best bandwidth efficiency.

 As shown in Fig. \ref{abastudy}(b), although the SC methods still perform better than conventional coding methods when SNR is low. The SR performance of the LC+SC is worse  than that of the SC. That means the compression of the LC reduces the redundancy of the message and increases the effect of semantic errors. The LC with traditional coding has a similar trend and loses its performance quickly as the SNR drop below the correction capability.

 According to the above discussion, the introduction of LC increases the compression capability and guaranties the performance in a proper compression ratio. In order to fully exert LLM functions, the key items require a slightly better SNR or extra protection strategy. This demonstrates the effectiveness of the proposed methods.

\section{Conclusion}
\label{s6}
This paper presents a communication framework for embodied AI agents that reduces transmission overhead while preserving task execution performance. Inspired by agentic AI workflows across distributed devices, the proposed approach exploits the content understanding, reasoning, and planning capabilities of LLMs. Specifically, LLM based compression is employed to shorten message length and effectively reduce redundancy in generated content. This mechanism is applicable to a wide range of generative agents. In addition, a semantic encoder-decoder is incorporated to enhance robustness over wireless links.

For task specific scenarios, the above mechanisms may introduce unavoidable information loss. In such cases, performance improvements rely on physical layer transmission strategies and knowledge base updates, where LLMs and semantic encoder-decoders exhibit complementary strengths. LLMs enable explicit semantic processing and can be readily enhanced through task specific knowledge and representative examples. Meanwhile, semantic encoder-decoders address channel variability through importance-aware transmission. Finally, ablation studies validate the effectiveness of the proposed design choices and provide insights into their respective contributions.

	\bibliographystyle{IEEEtran}
	\bibliography{bibtex0320}
	
	%
	
	
	
\end{document}